\documentstyle[sprocl]{article}

\input{psfig}

\def\Journal#1#2#3#4{{#1} {\bf #2}, #3 (#4)}

\def\be{\begin{equation}}
\def\ee{\end{equation}}
\def\bea{\begin{eqnarray}}
\def\eea{\end{eqnarray}}

\begin{document}

\title{ISOSPIN FROM SPIN BY COMPOSITENES\footnote{%
To appear in the proceedings of the workshop ``Lorentz
Group, CPT, and Neutrinos'' 26.--28. June 1999 Zacatecas/Mexico.}}

\author{Bertfried Fauser \& Heinz Dehnen}

\address{Universit\"at Konstanz\\Fakult\"at f\"ur Physik\\
Fach M678\\D-78457 Konstanz\\E-Mail:~Bertfried.Fauser@uni-konstanz.de\\
May 21, 1999} 

\maketitle\abstracts{ 
We propose a new method to generate the internal isospin degree of freedom
by non-local bound states. This can be seen as motivated by
Bargmann--Wigner like considerations, which originated from local spin
coupling. However, our approach is not of purely group theoretical origin,
but emerges from a geometrical model. The rotational part of the Lorentz
group can be seen to mutate into the internal iso-group under some
additional assumptions. The bound states can thereafter be characterized
by either a triple of spinors ($\xi_1, \xi_2$, $\eta$) or a pair of an
average spinor and a ``gauge'' transformation ($\phi, R$). Therefore,
this triple can be considered to be an isospinor. Inducing the whole dynamics
from the covariant gauge coupling we arrive at an isospin gauge theory and
its Lagrangian formulation. Clifford algebraic methods, especially the
Hestenes approach to the geometric meaning of spinors, are the most useful
concepts for such a development. The method is not restricted to isospin,
which served as an example only.
}

\section{Introduction}

Composite systems are of extraordinary importance in modelling
physical systems. Using composites as building blocks one can build up
more involved systems. However, an exact treatment of the bound systems
show up to be not only difficult but impossible in general. While two
body problems are usually solvable in classical mechanics, it has been
proved that this cannot be done in general already for a three body
configuration. Moreover, such a three body, or three degrees of
freedom, situation turns out to be in general chaotic. This might be called
the bound state problem. On the other hand, it is possible only to design
exact solvable models in very idealized situations which have but a few
contacts to real systems. 

Already Luis de Broglie suggested the photon to be 
composite.\cite{deB:54} Bargmann and Wigner \cite{BW:48} derived
locally coupled systems of spins. This story entered quantum field
theory via Heisenbergs unified non-linear spinor field theory,
\cite{H:67} which took the radical point to start with only one
fundamental spinor field considering all (other) observable fields as
composed quantities. A quite elaborate and extended version of a
non-linear spinor theory of principally unobservable preon
fields was developed by Stumpf, \cite{SB:94} a Heisenberg and
Bopp pupil, which succeeded in deriving, by exact calculations,
the electro-weak interaction and the corresponding bosons on a
quantum level as composite fields including their dynamics.
However, due to the complicated and quite involved
computations the strong and gravitational interactions resisted
up to now an exact treatment. Nevertheless, the Stumpf theory
of weak mapping can fairly well be applied to approximations,
when the $n$-particle wave functions are not known exactly.
This led to very fruitful calculations deriving QCD and a
linearized version of gravity and a successful application to
superconductivity including the important fact of symmetry
breaking. Our criticism of this method, especially of the
tedious composite wave function calculations and their
algebraic ill-defined behaviour, was formulated in.\cite{FS:97}

The method proposed in this paper relays, as the Stumpf theory, on
an extension of the Bargmann--Wigner idea of local spin
coupling. Since Stumpf tried to calculate non-local bound
state wave functions as exact solutions of Bethe--Salpeter like
equations, we surmount this difficulty by a simple geometric
postulate. Hence, no dynamic origin of the bound mechanism is
given.

Beside the fact, that also in the Bargmann--Wigner local
coupling no such mechanism was proposed, we will see later,
that this step puts the bound state problem in a nutshell and
opens thereby the study of bound-state dynamics even if the
details are not known exactly. This is much more important in
non-abelian gauge theories like ${\bf SU}(2)$ electro-weak and
${\bf SU}(3)$ colour of QCD which are chaotic in the classical
formulation.

\section{Geometry of Hestenes spinors}

The main idea, proposed by Hestenes\cite{H:66} since 1966, is to
formulate Dirac theory representation independent. In usual, old
fashioned notation, e.g. Bjorken and Drell, \cite{BD:66} Diracs 
equations for a free field are written as
\bea
i\gamma_{\alpha\beta}^\mu \partial_\mu \psi_\beta
-m\psi_\alpha &=& 0.
\eea
The common interpretation is to look at the $\gamma_{\alpha\beta}^\mu$
matrices as {\em matrix valued components}\/ of a four-vector
$\vec{\gamma}$. With this preconception one runs immediately into
difficulties. The spin indices $\alpha, \beta$ cannot be considered to
characterize components of the same space ${\bf M}_{1,3}$ (Minkowski
space) as the $\mu$ index does. A so-called ``internal'' spin space, no
comment what ``internal'' does mean here or is related to, is introduced.
Denoting spin space ${\cal S}$, one has $\vec{\gamma}$ to
be an element of a mixed tensor space
\bea
\vec{\gamma} \,=\,
\gamma^\mu_{\alpha\beta} \, e_\mu\otimes\xi_\alpha\otimes\xi_\beta
&\in& 
{\bf M}_{1,3} \otimes {\cal S} \otimes {\cal S}^*
\cong < \gamma_{\alpha\beta}^\mu >.
\eea
Using Ockhams razor, Hestenes simply identified the $\gamma$-matrices
with the spinor representation matrices of the Minkowski basis vectors.
Denoting the half integral spinor representations as $D^{\frac{1}{2}0}$
and $D^{0\frac{1}{2}}$, we can give a representation $\pi$ of an
arbitrary but fixed basis $\{ e^\mu\} \in {\bf M}_{1,3}$ as spin-tensors
\bea
\pi(e^\mu) &=& \gamma^\mu \in D^{\frac{1}{2}0} \otimes D^{0\frac{1}{2}}
\,=\, \gamma_{\alpha\beta}^\mu \in {\cal S}_\alpha \otimes {\cal
S}^*_\beta.
\eea
Here $\mu$ is a label, {\em not}\/ an index! Observe that these
representation spaces {\em include}\/ the Minkowski space. In this way,
no
additional internal space is needed.

Since the $\gamma$-matrices constitute a Clifford algebra over the reals
${\bf R}$, which can be denoted ${\bf R}_{1,3}$, one has the following
decomposition
\bea
{\bf R}_{1,3} &=& 1 \oplus {\bf M}_{1,3} \oplus B \oplus T \oplus V 
\,=\, {\cal S} \otimes {\cal S}^*,
\eea
where $B={\bf M}_{1,3}\wedge {\bf M}_{1,3}$ is the space of bi-vectors,
$T$ are tri-vectors and $V$ is the four-vector volume element. A
complexification can be achieved as ${\bf C} \otimes {\bf R}_{1,3} \cong
{\bf C}_{1,3}$ if desired. From this decomposition it is clear that the
multivector spaces are invariant subspaces of the Clifford-Lipschitz group
and especially of the various pin and spin groups, e.g.:
${\bf spin}_{1,3} \bullet {\bf M}_r \subset {\bf M}_r.$
Turning Diracs equations into a representation free scheme, yields the
Dirac-Hestenes equation in terms of an arbitrary basis $\{ e_\mu \}$ of
the Minkowski space
\bea
\partial \Psi e_{12} - m \Psi e_0 &=& 0.
\eea
Notations are: $e_{12}=e_1e_2$, $\partial = e^\mu\partial_\mu$ and $\Psi
\in {\bf R}^+_{1,3}$ an even operator spinor. The condition to be even
reduces the number of degrees of freedom to the correct value and splits
the spin-tensor into two parts
$
{\cal S} \otimes {\cal S} = 
({\cal S} \otimes {\cal S})^+ \oplus
({\cal S} \otimes {\cal S})^-,
$
where only the even part is again a subalgebra. However since one is
interested in  this spaces as {\em representation spaces},\/ only the
linear structure is in use. The equation respects this decomposition if
and only if the mass term vanishes, see.\cite{F:96} Introducing
$\Psi^\prime = \Psi e_{12}$, we arrive at the equation
$\partial \Psi^\prime = 0.$
 
The important thing is, that $\Psi$ as an algebra element obtains an
operational interpretation.  For $1$, $i_4$ or $\gamma^0$ one gets
\be
\Psi 1 \Psi\tilde{~} = \Omega_1,\quad
\Psi i_4 \Psi\tilde{~} = \Omega_2 i_4,\quad
J^\mu = <\Psi \gamma^0 \Psi\tilde{~}>_1 \,=\,
< \Psi \gamma^0 \Psi\tilde{~}\gamma_\mu>_0\gamma^\mu,
\ee
where $\tilde{~}$ is the reversion antiautomorphism and $<\ldots>_1$ the
projection to the one-vector part.  We can give a polar decomposition of
$\Psi$:
\bea
\Psi &=&
 (\Psi \Psi\tilde{~})^{\frac{1}{2}}
\frac{\Psi}{ 
 (\Psi \Psi\tilde{~})^{\frac{1}{2}} } 
= \rho^\frac{1}{2} V,
\eea
where $V$ is in the Clifford-Lipschitz group. Since spinors are defined
only up to a phase, we extract the duality rotation with Yvon-Takabayasi
angle $\beta$ from $V$, which Hestenes interprets as 
statistical.\cite{HG:75,GH:75} The polar form is then
\bea
\Psi &=& \rho^{\frac{1}{2}}V \,=\,
\rho^{\frac{1}{2}}e^{i_4\frac{\beta}{2}} R.
\eea
$R$ has six degrees of freedom (eight of $\Psi$
minus $\rho, \beta$) and is an element of the double cover of the Lorentz
group ${\bf spin}_{1,3}$, inducing Lorentz transformations on vectors. A
Dirac-Hestenes spinor is thus an operational object, which transforms
a reference object into locally given ones. As an example
\be
J^\mu(x)\,=\, \Psi(x) e^\mu\Psi\tilde{~}(x) 
\,=\,
\rho(x) e^{i_4\beta(x)} R(x)e^\mu R\tilde{~}(x) 
\,=\, \rho(x) e^{i_4\beta(x)} L^\mu_\nu(x)e^\nu.
\ee
The current density is seen to be $\rho(x)$ times a duality rotation
times a local Lorentz transformation $L^\mu_\nu(x)$ of the fixed reference
system $\{ e^\mu \}$. This interpretation will be the starting point for
our geometric composite model.

\section{Geometric model for bound states}

\subsection{Bargmann-Wigner bound states}

Recalling the idea of Bargmann and Wigner, we start with
free spinors. How are dynamical equations derived for spin tensors?
Restricting ourself to two particle systems, every rank two spin
tensor can be decomposed into irreducible parts using the symmetric group.
We obtain
\bea
\xi_1 \otimes \xi_2 &=&
\xi_{(1} \otimes \xi_{2)} \oplus 
\xi_{[1} \otimes \xi_{2]}
\eea
where $[12]$ and $(12)$ means (anti) symmetrization of the tensors. Since
the $\xi_i$ are supposed to cary a $D^{\frac{1}{2}}$ spin one-half
representation, we end up with a spin one triplet (symmetric case) and a
spin zero singlet (antisymmetric case). Concentrating on the symmetric
case, we can give the equations of motion as\cite{L:68}
\bea
(\gamma^\mu\partial_\mu -m)_{\alpha\alpha^\prime}
\Psi_{(\alpha^\prime\beta)} &=& 0.
\eea
Expanding this in a symmetric basis of $\gamma$-matrices $
\gamma^\mu C$,
$
\Sigma^{\mu\nu}C= 
\frac{1}{2}(\gamma^\mu\gamma^\nu-\gamma^\nu\gamma^\mu)C
$,
where $
C\gamma^\mu C^{-1} \,=\, -\gamma^{\mu T},
$ leads to a decomposition of $\Psi$ into
\bea
\Psi_{(\alpha\beta)} &=&
A_\mu\,(\gamma^\mu C)_{\alpha\beta}
+F_{\mu\nu}\,(\Sigma^{\mu\nu}C)_{\alpha\beta}.
\eea
Since $A_\mu$ has four and $F_{\mu\nu}$ has six components, they cannot be
independent. One finds the abelian ${\bf U}(1)$ gauge theory 
\bea
&F_{\mu\nu} \,=\, \partial_\mu\,A_\nu - \partial_\nu\,A_\mu
\,\cong\, \mbox{d}A& \\ \nonumber
&\Box A_\mu-\partial_\mu(\partial_\nu A_\nu) \,=\, m^2 A_\mu\,.&
\eea
In a suitable gauge and the limit of vanishing mass one arrives at the
theory of vacuum electrodynamics. Spin zero particles or spin $3/2$ Proca
fields emerging from third rank spin-tensors can be treated analogously.

However, it can be shown, that e.g. gravitational forces cannot be modelled
in this way. Also several other difficulties remain unsolved. Finally we
remark, that there is no direct contact between the dynamics of the
constituents and the composite. Bargmann and Wigner assume simply the same
differential operator to act on every tensor of any rank.

\subsection{Non-local geometric bound states}

A geometric composite model is proposed, which is based on the
Hestenes interpretation of Dirac spinors. This model is also {\em ad
hoc}\/ in some sense, but cures some of the above mentioned problems of
local Bargmann-Wigner states.

Given two Dirac particles $\xi_1$ and $\xi_2$ forming a non-local bound
system, we do not concentrate on informations about the
internal dynamics, but try to give a dynamics for the entire bound system. 
We propose:
i) The bound object shall be charaterizable by the usual quantum
numbers.\\
ii) The dynamics of composites shall follow from the dynamics of the free
constituents.\\
From our consideration we conclude the following model:
Let ($\xi_1,\xi_2,\eta$) be a triple of spinors. The $\xi_i$ are the
spinors of the constituents and $\eta$ is a reference spinor connected
to the $\xi_i$ by $\xi_i = \rho_i^{1/2}exp(i_4\beta/2)R_i \,\eta$.
Assume that there is a very small vector $\vec{a}$ satisfying $\vec{x}_2 =
2*\vec{a}+\vec{x}_1$. The time average over an internal period of
$\vec{a}$ shall be constant. Since the $\xi_i$ spins are space-like
separated, they can have {\em any alignment}.\/ However we will discard
here reflections, boosts and assume that rotations can map the spinors
onto one another. We introduce hence an average spinor $\phi$ and a
Lorentz transformation $R$ which describes all internal degrees of freedom
of the composite beside internal radial oscillations, boosts and mutual
internal reflections. In fact this removes an internal phase. One
observes the correspondence
\bea
(\xi_1,\xi_2,\eta) &\cong& (\phi,R) \\ \nonumber
\#\, 7+7+0 &=& 7+3+(3\mbox{~boosts}+1\mbox{~radial}),
\eea
where the second line gives the degrees of freedom. The reference spinor
$\eta$ is not subjected to change, and has no freedom. 

\begin{figure}
\centerline{\psfig{figure=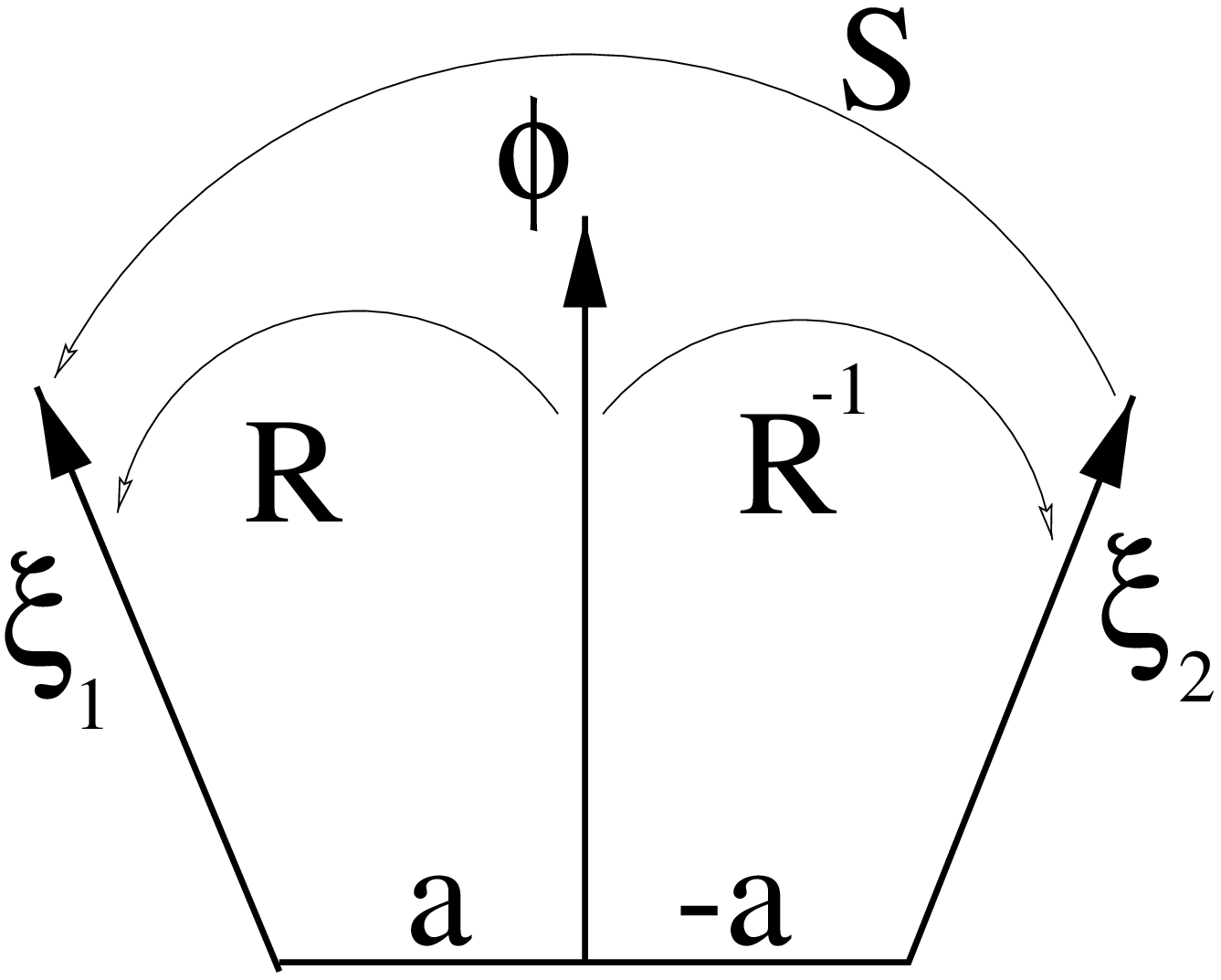,height=3.0cm,width=3.0cm}}
Figure 1: Geometrical model of a non-local bound object described by an
average spinor $\phi$ and a Lorentz rotation $R$ {\em or}\/ two
constituents $\xi_i$.
\end{figure}

\section{Induced gauge theory}

We derive the covariant coupling of $\phi$ and $R$ as a direct 
consequence of our geometric composite model. Thereafter the
theory is completed by inducing the full action from the covariant
coupling. The curvature two-form or kinetic field energy term is added to
turn the local Lorentz field $R$ into a dynamical one, which has
consequently to be chosen to be a boson.

{}From our geometrical model we derive that there exits a Lorentz
transformation $S$ which connects the two spinors $\xi_i$, see Figure 1.:
\be
S\xi_2 = \xi_1,\quad
S^{-1}\xi_1 = \xi_2,\quad
S\tilde{~} = S^{-1}.
\ee
The inverse exists because of the group structure. But observe, that we
can also build the square root if we restrict ourselfs to the compact part
of the Lorentz group (rotations). More generally this can be done if the
spin directions are not on the light-cone, which is impossible here.
Hence
we define
\be
R^2=S,\quad
R\tilde{~}=R^{-1}.
\ee
Introduce a centre of mass or average spinor $\phi$ which
satisfies the following two equations (which actually defines $\phi$)
\be
\xi_1 = R\phi,\quad
\xi_2= R^{-1}\phi.
\ee
More precisely we should include the translations $T_{\vec{a}}$ also,
but in a dilute gas at moderate temperatures the particles will usually be
far away from another and lock point like, so internal vibrations are not
important. 

In deriving the dynamics we have to remember that $R(x)$ is a field.
Furthermore, we use the free field dynamics for the $\xi_i$ to derive the
composite dynamics. Starting from the Dirac-Hestenes equation
\be
\nabla \xi_1 e_{12} + m\, \xi_1 e_0 =0
\ee
inserting $\xi_1=R\phi$, we get
\be
\nabla (R\phi) e_{12} + m (R\phi) e_0 =0.
\ee
Using $RR\tilde{~}=1$ and the Leibnitz rule yields $
(R\tilde{~}(\nabla R)+ \nabla) \phi e_{12} + m\phi e_0 =0 $
which can be summarized with help of the {\em covariant derivative}\/ 
$D:= \nabla + R\tilde{~}(\nabla R)$ to
\be
D\phi e_{12} + m\phi e_0 =0.
\ee
Indeed, $R\tilde{~}(\nabla R)$ can be considered to be a vector field.
Following Hestenes \cite{H:86} we might introduce
$\Omega = 2R\tilde{~}(\nabla R)$ to obtain the derivative of $R$ by
$\nabla R = R/2(2R\tilde{~}(\nabla R))=1/2 \,R\Omega$. The covariant
derivative reads now
\be
D = \nabla + \frac{1}{2} \Omega,
\ee
where the coupling constant is included in $\Omega = g\,\Omega^\prime$.
The field strength can easily be calculated as usual by computing the
commutator of covariant derivatives, which yields a non-abelian ${\bf
SU}(2)$ from the covering group of the rotational part of the Lorentz
group,
\be
F \,=\, [D,D] \,=\, [\nabla+\frac{1}{2}\Omega ,
\nabla+\frac{1}{2}\Omega ].
\ee
In a symbolic notation this can be written as
\be
F \,=\, \mbox{d} \Omega + \frac{1}{4} \Omega \times \Omega
\ee
where the first term is structural equivalent to the abelian field
strength and the second term is the non-abelian left regular action of
${\bf spin}_{3}\cong{\bf SU}(2)$ on the Lie group generators. This sort of
action can now be seen {\em to constitute an internal isospin}\/ of the
geometric composite proposed above. The full action of this theory can be
written as
\be
{\cal L} \,=\,
<\phi (D + \frac{1}{2}\Omega )\phi\tilde{~} + 
  \frac{1}{4} F^2 >_0\,.
\ee 
It occurs now, that from a geometrical composite model, using operational
spinors, we were able to derive a ${\bf SU}(2)$ gauge covariant coupling
and a ${\bf SU}(2)$ gauge theory by inducing the gauge field dynamics.
The
{\em reversed}\/ statement would be, that every gauge field might be seen
as an internal motion. Since only compact groups are used, this internal
degrees of freedom are due to bound structures. Our approach should be
compared with the purely gauge theoretic treatment of Daviau,\cite{D:93}
which shows clearly the isospin acting on different representations.

Remark that for a single spinor it is {\em not}\/ necessary to introduce a
reference spinor $\eta$, while this is un-evitable, due to relative
adjustments, for two or n-particle systems. It is possible to account for
internal radial oscillatory motions in our approach, which then would
``gauge'' $T_{\vec{a}}$.

Since we omitted boosts, which are however not compact, our bound state
concept is not relativistic invariant, in-spite of the resulting theory
which is. This has to be considered elsewhere.

The geometric bound model is in accord with some very general
considerations on linear forms of multiparticle systems which were
investigated in.\cite{F:99} It furthermore explains a factor 2 not
understood in the Stumpf weak mapping formalism which there forced the
temporal gauge to be used.\cite{P:90,SB:94}

Finally a strong support is given, that gauge theories might arise
always from dynamical effects in theories of {\em compound objects}.\/ 

\section*{Acknowledgement}

The first author, B.F.,was partly supported by the DFG.


\section*{References}
\begin{thebibliography}{99}

\bibitem{deB:54} 
L. de Broglie,
{Th\'eorie g\'eneral des particules \'a spin}
{Gautier--Villars/Paris,2nd. ed.}
{(1954)} 

\bibitem{BW:48} 
V. Bargmann, E. Wigner,
\Journal{Proc. Nat. Acad. Sci. (USA)}{34}{211--223}{1948}

\bibitem{H:67} 
W. Heisenberg, 
{Einf\"uhrung in die einheitliche Feldtheorie der Elementarteilchen}
{Hirzel/Stuttgart} 
{(1967)}

\bibitem{SB:94} 
H. Stumpf, Th. Borne, 
{Composite particle theory in quantum field theory}
{Vieweg/Braunschweig}
{(1994)}

\bibitem{FS:97} 
B. Fauser, H. Stumpf,
\Journal{Adv. in Appl. Cliff. Alg.}{7(sup.)}{399--418}{1994}

\bibitem{H:66} 
D. Hestenes,
{Space time algebra}
{Gordon and Beach}
{(1966)}

\bibitem{BD:66} 
J.D. Bjorken, S.D. Drell,
{Relativistische Quantenmechanik}
{Mc Graw-Hill inc. 1964, BI-Wissenschaftsverlag/Mannheim}
{(1966)}

\bibitem{F:96} 
B. Fauser,
\Journal{Proc. of the 4th Int. Conf. on Clifford alg. and their appl. in
Math. Phys., Aachen, Kluwer/Dordrecht}{}{89--107}
{1996}

\bibitem{HG:75} 
D. Hestenes, R. Gurtler,
\Journal{J. Math. Phys.}{16(3)}{556}{1975}

\bibitem{GH:75} 
R. Gurtler, D. Hestenes,
\Journal{J. Math. Phys.}{16(3)}{573}{1975}

\bibitem{L:68}
D. Lurie,
{Particle and Fields}
{Interscience Publ.}
{(1968)}

\bibitem{H:86}
D. Hestenes,
{New foundation for classical mechanics}
{Kluwer/Dorchrecht}
{(1986)}

\bibitem{D:93}
C. Daviau,
{Equation de Dirac non lineaire,}
{Thesis, Univ. Nantes}
{(1993)}

\bibitem{F:99} 
B. Fauser, {Clifford geometric parameterization of inequivalent vacua,}
\Journal{J. Phys. A: Math. Gen.}{}{submitted}{1999} hep-th/9710047

\bibitem{P:90}
W. Pfister,
{Yang-Mills-Dynamik als effektive Theorie von vektoriellen
Spinor-Isospinor-Bindungszust\"anden in einem Preonfeldmodell,}
{Thesis, Univ. T\"ubingen}
{(1990)}
\end{thebibliography}
\end{document}